\documentclass[oneside,a4paper,reqno]{amsart}
\usepackage{amssymb}
\usepackage{graphics}
\usepackage{bm}

\begin{document}
\begin{titlepage}
\begin{center}
\bfseries  	THE HESS-PHILIPP MODEL IS NON-LOCAL
\end{center}
\vspace{1 cm}
\begin{center} D M APPLEBY
\end{center}
\begin{center} Department of Physics, Queen Mary,
University of London,  Mile End Rd, London E1 4NS,
UK
 \end{center}
\vspace{0.5 cm}
\begin{center}
  (E-mail:  D.M.Appleby@qmul.ac.uk)
\end{center}
\vspace{0.75 cm}
\vspace{1.25 cm}
\begin{center}
\textbf{Abstract}\\
\vspace{0.35 cm}
\parbox{10.5 cm }{ 
Hess and Philipp have constructed what, they claim, is a local hidden variables
model reproducing the empirical predictions of quantum mechanics.  In this paper
explicit expressions for the  conditional probabilities for the outcomes of the
measurements at the two detectors are calculated.  These expressions provide a
conclusive demonstration of the falsity of the authors' claim.  The authors
 give two different accounts of their model.  The published version 
omits a crucial detail.  As a result it disagrees with quantum mechanics. 
It also violates signal locality.  The unpublished version agrees with quantum
mechanics.  However, it violates the condition of parameter independence, as
Myrvold has previously shown.}
\end{center}
\end{titlepage}
\section{Introduction}
Hess and Philipp~\cite{Hess1,Hess2} (also see Hess and Philipp~\cite{Hess3,Hess4})
have constructed what, they claim, is  a local hidden-variables model reproducing
the quantum mechanical predictions for the singlet state---contrary to the result
proved by Bell~\cite{Bell}.  The falsity of their claim has been shown by,
among others, 
Myrvold~\cite{Myrvold} (also see Gill \emph{et
al}~\cite{Gill1,Gill2} and
Mermin~\cite{Mermin}).  However, Hess and Philipp~\cite{Hess5} dispute Myrvold's
conclusion. The purpose of this note is to present some additional considerations
which, we hope, may help to settle the question.

Hess and Philipp desribe their model in two different places.  The two versions
are not identical.   By some oversight the published version~\cite{Hess1} 
(version 1) omits a crucial detail.  As a result it disagrees with quantum
mechanics.  It also violates
\emph{signal} locality.  The unpublished  version~\cite{Hess2} of their
model (version~2) does not suffer from these deficiencies.  However, the same
equations which show that version~1 violates signal locality also conclusively
demonstrate that version~2 violates parameter independence, as Myrvold has argued.

The concept of parameter independence was  analyzed in detail by
Jarrett~\cite{Jarrett} and Shimony~\cite{Shimony1,Shimony2}.  Consider a pair of
spin-$1/2$ particles prepared in the singlet state.  Particle $1$  is sent to
station $S_1$, which measures its spin in the direction $\mathbf{a}$, obtaining
the outcome $A=\pm 1$.   Particle $2$  is sent to station $S_2$, which measures
its spin in the direction $\mathbf{b}$, obtaining the outcome $B=\pm 1$.  Let
$\lambda$ denote  the complete hidden state of the pair $1+2$. 
Let $p_1 (A|\mathbf{a}, \mathbf{b}, \lambda)$ be the
 probability of obtaining outcome $A$ at $S_1$  for
detector settings 
$\mathbf{a}$, $\mathbf{b}$ and hidden state   $\lambda$.
Similarly, 
let $p_2 (B|\mathbf{a}, \mathbf{b}, \lambda)$ be the
 probability of obtaining outcome $B$ at $S_2$  for given
$\mathbf{a}$, $\mathbf{b}$, $\lambda$.  

A hidden variables model satisfies the condition of parameter
independence (``locality'', in the terminology of Jarrett~\cite{Jarrett}) if and
only if
$p_1$ is independent of
$\mathbf{b}$ and $p_2$ is independent of $\mathbf{a}$, so that 
\begin{align}
p_1 (A|\mathbf{a}, \mathbf{b}, \lambda)
& = p_1 (A|\mathbf{a}, \lambda)
\label{eq:ParIndA}
\\
\intertext{and}
p(B|\mathbf{a}, \mathbf{b}, \lambda)
& = p(B|\mathbf{b}, \lambda)
\label{eq:ParIndB}
\end{align}
for suitable 
$p_1(A|\mathbf{a},\lambda)$ and $p_2(B|\mathbf{b}, \lambda)$.  

Version~2 of the Hess-Philipp model~\cite{Hess2} violates parameter
independence.  This clearly implies that the model is non-local because
it means that if, \emph{per impossibile}, one could prepare an ensemble
of pairs all in the same hidden state
$\lambda$, then experimenters at
$S_1$ and
$S_2$ could use these pairs to communicate superluminally.

This point has already been made by Myrvold~\cite{Myrvold}.  However, Myrvold's
paper is largely concerned with a simplified version of the Hess-Philipp model. 
The extension to the full model is only made briefly, at the end of his paper.  In
particular,  Myrvold  does not actually calculate the conditional probabilities
appearing in Eqs.~(\ref{eq:ParIndA}) and~(\ref{eq:ParIndB}).  This gives Hess and
Philipp~\cite{Hess5} some scope to  challenge his conclusion.  

In the following we do calculate the conditional probabilities, and
explicitly show that they do not satisfy Eqs.~(\ref{eq:ParIndA})
and~(\ref{eq:ParIndB})---thereby providing a clear-cut mathematical
demonstration that Myrvold is correct.  Hess and Philipp are only
entitled to disagree with this statement if they can identify a flaw in
our
\emph{calculations}, and if they can present some alternative, fully
explicit calculations leading to expressions which \emph{do} satisfy  
Eqs.~(\ref{eq:ParIndA})
and~(\ref{eq:ParIndB}).

In the following we begin by showing that version~1 of the Hess-Philipp model
violates signal locality.  We  then go on to show that, for what is
essentially the same reason, version~2 violates the condition of parameter
independence. 

\section{Version~1 of the Model Violates Signal Locality}

Hess and Philipp construct their model in two stages.  The first stage is
the theorem proved on pp.~14231--2 of ref.~\cite{Hess1}.  In Eqs.~(22),
(23) and (27)  they define functions
$A_{\mathbf{a}}(u)$, 
$B_{\mathbf{b}}(v)$ and a density $\rho_{\mathbf{a} \mathbf{b}}(u,v)$
with the property 
\begin{equation}
\int A_{\mathbf{a}}(u) B_{\mathbf{b}}(v)
\rho_{\mathbf{a}\mathbf{b}} (u,v) du dv = - \mathbf{a}\cdot \mathbf{b}
\end{equation}
(Eq.~(18) in the statement of their theorem).  The reader may confirm
that one also 
has\footnote{
The reader who does wish to check this statement should bear in mind that there is a
missing summation sign in Eq.~(26) of ref.~\cite{Hess1}, as noted in Hess
and Philipp~\cite{Hess5}}
\begin{align}
\int A_{\mathbf{a}}(u) 
\rho_{\mathbf{a}\mathbf{b}} (u,v) du dv
& = \mathbf{a} \cdot |\mathbf{b}| + 
\frac{1}{2}
\sum_{k=1}^{3} \sum_{r=1}^{n/2} N_{2r} (|a_k|) \psi_{2r} (|b_k|)
\label{eq:ExpA1}
\\
\int B_{\mathbf{b}}(u) 
\rho_{\mathbf{a}\mathbf{b}} (u,v) du dv
& = -  |\mathbf{a}| \cdot \mathbf{b} 
\label{eq:ExpB1}
\end{align}
Here $|\mathbf{a}|$, $|\mathbf{b}|$ are the vectors with components 
$|a_1|, |a_2|, |a_3|$ and $|b_1|, |b_2|, |b_3|$ respectively.  
$N_{2 r}$, $\psi_{2 r}$ and the even integer $n$
are the quantities defined in the lemma on p.~14231 of
ref.~\cite{Hess1}.

The second stage in the  construction of the Hess-Philipp model is the
complicated combinatoric 
argument
leading to
Eq.~(35) of ref.~\cite{Hess1}.  It should be noted that there are two
problems with this part of the construction, one major and one minor.

The major problem is the point made by  
  Gill
 \emph{et al}~\cite{Gill2}, that the 
  index $m$ featuring in the construction apparently introduces an element
of
  non-locality into the model.  In our view
  Hess and Philipp~\cite{Hess5} do not satisfactorily answer this
  objection.   However, we will not
   insist on
  the point here because it is tangential to our argument.

The minor problem  is a technical point, concerning the details
of the combinatorics.  We discuss it in the appendix.

The full probability
space for the Hess-Philipp model consists of all pairs $(\lambda,
\omega)$, where
$\lambda$ describes the state of the source particles and $\omega$ refers
to  the detectors (second paragraph on p.~14230 of
ref.~\cite{Hess1}).  Let
$\nu_{\omega}$ be the probability measure on the space of $\omega$, and
let
$\nu_{\lambda}$ be the probability measure on the space of $\lambda$. 
For any random variable $X$, let 
$E_{\lambda}(X) =
\int X \, d
\nu_{\omega}$ be the conditional expectation value obtained by integrating
out
$\omega$ for a fixed value of
$\lambda$. 

The combinatorics leading to Eq.~(35) of ref.~\cite{Hess1} are such that

\begin{equation}
E_{\lambda} (A_{\mathbf{a}}  B_{\mathbf{b}})
  = \int A_{\mathbf{a}}(u) B_{\mathbf{b}} (v)
\rho_{\mathbf{a}\mathbf{b}} (u,v) du dv  = 
  - \mathbf{a}\cdot \mathbf{b} 
\label{eq:ABCondExp}
\end{equation}
holds trivially.   It is also
easily seen that
\begin{equation}
E_{\lambda}(B_{\mathbf{b}})  =
\int B_{\mathbf{b}}(u) 
\rho_{\mathbf{a}\mathbf{b}} (u,v) du dv
 = - |\mathbf{a}|\cdot \mathbf{b} 
\label{eq:BCondExp}
\end{equation}
However, the evaluation of $E_{\lambda}(A_{\mathbf{a}})$ is
slightly less straightforward.  It is shown in the appendix that
\begin{equation}
E_{\lambda}(A_{\mathbf{a}})  =
\mathbf{a}\cdot |\mathbf{b}| + \frac{1}{2} \left(1 -
|\mathbf{a}| \cdot |\mathbf{b}|\right) + \frac{\theta}{16 n^2} 
\label{eq:ACondExp}
\end{equation}
with $0 \le \theta \le 1$.

Now let $E(X) = \int E_{\lambda} (X) \, d \nu_{\lambda}$ be the
unconditioned expectation value of $X$. A striking 
feature\footnote{
It is a curious feature because it means that all the work 
is being done by the detectors.  The source particles might as well
not be there.
} of
Eqs.~(\ref{eq:ABCondExp}--\ref{eq:ACondExp}) is that the right-hand sides
are independent of $\lambda$.  Consequently, the unconditioned
expectation values are the same as the conditional ones:
\begin{align}
E (A_{\mathbf{a}}  B_{\mathbf{b}})
 & = 
  - \mathbf{a}\cdot \mathbf{b} 
\label{eq:ExpAB}
\\
E(B_{\mathbf{b}}) & = 
 - |\mathbf{a}|\cdot \mathbf{b} 
\label{eq:ExpB}
\\
E(A_{\mathbf{a}}) & = 
\mathbf{a}\cdot |\mathbf{b}| + \frac{1}{2} \left(1 -
|\mathbf{a}| \cdot |\mathbf{b}|\right) + \frac{\theta}{16 n^2}
\label{eq:ExpA} 
\end{align}
whatever the probability measure $\nu_{\lambda}$.

At this stage we notice that the model disagrees with quantum mechanics. 
For the singlet state 
 quantum mechanics predicts  
$E(A_{\mathbf{a}})=E(B_{\mathbf{b}})=0$ for all $\mathbf{a}, \mathbf{b}$. 
It can be seen that Eqs.~(\ref{eq:ExpB}) and~(\ref{eq:ExpA}) disagree with
this prediction.  

Not only does the model disagree with
quantum mechanics.  It also violates \emph{signal} 
locality\footnote{
We are indebted to M.~\.{Z}ukowski for this observation.
}.  To see this, let $p_2 (B=\pm 1 |\mathbf{a}, \mathbf{b})$ be the
probability of obtaining the measurement outcome $B=\pm 1$ at station
$S_2$, for detector settings $\mathbf{a}, \mathbf{b}$.  Then
\begin{equation}
p_2 (B=\pm 1 |\mathbf{a}, \mathbf{b}) = \frac{1}{2} 
\bigl(1\pm E(B_{\mathbf{b}})\bigr)
=
\frac{1}{2} (1 \mp |\mathbf{a}| \cdot \mathbf{b})
\label{eq:ProbBVer1}
\end{equation}
Suppose that Alice at station $S_1$ and Bob at station
$S_2$ have previously agreed that Bob will always measure in the
direction $(-1,0,0)$, while Alice will measure in one of the two
alternative directions $(1,0,0)$ or $(0,1,0)$.  Then, if Alice measures
in the direction $(1,0,0)$ there will be probability 1 of Bob obtaining
the result
$+1$, while if Alice measures in the direction
$(0,1,0)$ there will only be probability $1/2$ of Bob obtaining the
result $+1$.  In this way they can use the arrangement to communicate
superluminally, with a probability of error that, with sufficient
redundancy, can be made arbitrarily small.

\section{Version~2 of the Model Violates Parameter Independence}

We now turn to version~2 of the Hess-Philipp model, described in
the e-print~\cite{Hess2}. Unlike version~1, version~2 of the model does reproduce
the empirical predictions of quantum mechanics (for the singlet state), and it
does not violate signal locality.   Nevertheless, it is still non-local. 
Furthermore, it is non-local for a reason that is closely related to the reason
that version~1 violates signal locality---as we now show.

The  crucial detail, which is omitted from version~1 of the model, is described 
at the end of ref.~\cite{Hess2}, in Section~5.3.  Let $A_{\mathbf{a}} (\lambda,
\omega)$,
$B_{\mathbf{b}} (\lambda, \omega)$ be the functions describing the measurement
outcomes in version~1 of the model.  Version~2 is obtained by making the
replacements
\begin{align}
A_{\mathbf{a}} (\lambda, \omega) & \to
r(\lambda) A_{\mathbf{a}} (\lambda, \omega)
\\
B_{\mathbf{b}} (\lambda, \omega) & \to
r(\lambda) B_{\mathbf{b}} (\lambda, \omega)
\end{align}
where 
$r(\lambda)$ is a function taking the values $\pm 1$, and having
 the property
$\int r(\lambda) d\nu_{\lambda}=0$ ($\nu_{\lambda}$ being the probability
measure on the space of the source variables $\lambda$, as 
before)\footnote{
   Hess and Philipp stipulate that $r$ depends on any ``parameter
   specific to the source (e.g. $\lambda_1$, $\lambda_2$ or time $t$)''.
   By ``time $t$'' they presumably mean the time at which the particles are
   emitted by the source.  In the definition of parameter independence,
   as given in Eqs.~(\ref{eq:ParIndA}), (\ref{eq:ParIndB}) above,
   $\lambda$ denotes a \emph{complete} specification of the 
   state of the source.  On this defintion $\lambda$ must
   be taken to  include
		 a specification of the time of emission. 
}.

With these replacements the model does reproduce the correct quantum mechanical
predictions for the unconditioned expectation values 
$E(A_{\mathbf{a}} B_{\mathbf{b}})$, $E(A_{\mathbf{a}})$ ,
$E(B_{\mathbf{b}})$ (in the singlet state).  Consequently, it does not violate
signal locality.

Suppose, however, that one considers the \emph{conditional} expectation values
 $E_{\lambda}(A_{\mathbf{a}})$,
$E_{\lambda}(B_{\mathbf{b}})$.  Then one has, in place of 
Eqs.~(\ref{eq:BCondExp}), (\ref{eq:ACondExp}) above,
\begin{align}
E_{\lambda}(A_{\mathbf{a}}) & =
r(\lambda) \left(\mathbf{a}\cdot |\mathbf{b}| + \frac{1}{2} \left(1 -
|\mathbf{a}| \cdot |\mathbf{b}|\right) + \frac{\theta}{16 n^2}\right)
\label{eq:ACondExp2}
\\
E_{\lambda}(B_{\mathbf{b}})  
& = - r(\lambda) |\mathbf{a}|\cdot \mathbf{b}
\label{eq:BCondExp2}
\end{align}
Let $p_1$, $p_2$ be the probabilities appearing on the left-hand sides
of Eqs.~(\ref{eq:ParIndA}), (\ref{eq:ParIndB}) above.  Then
\begin{align}
p_1(A =\pm 1|\mathbf{a}, \mathbf{b}, \lambda) & =
\frac{1}{2}\left(1 \pm E_{\lambda }(A_{\mathbf{a}}) \right)
\\
p_2(B =\pm 1|\mathbf{a}, \mathbf{b}, \lambda) & =
\frac{1}{2}\left(1 \pm E_{\lambda }(B_{\mathbf{b}}) \right)
\end{align}
implying
\begin{align}
p_1(A =\pm 1|\mathbf{a}, \mathbf{b}, \lambda) & =
\frac{1}{2}\left(1 \pm 
r(\lambda) \left(\mathbf{a}\cdot |\mathbf{b}| + \frac{1}{2} \left(1 -
|\mathbf{a}| \cdot |\mathbf{b}|\right) + \frac{\theta}{16 n^2}\right)\right)
\\
p_2(B =\pm 1|\mathbf{a}, \mathbf{b}, \lambda) & =
\frac{1}{2}\bigl(1 \mp   r(\lambda) |\mathbf{a}|\cdot \mathbf{b} \bigr)
\label{eq:ProbBVer2}
\end{align}
These expressions clearly fail to satisfy the condition of parameter
independence,  stated in
Eqs.~(\ref{eq:ParIndA}), (\ref{eq:ParIndB}) above. 
 It follows that version~2 of the model is
non-local.

The violation of parameter independence in version~2 of the model is closely
related to the violation of signal locality in version~1.  This can be seen by
comparing the expression for 
$p_2(B=\pm 1| \mathbf{a}, \mathbf{b},\lambda)$ in version~2 (see 
Eq.~(\ref{eq:ProbBVer2}) above) with the expression for 
$p_2(B=\pm 1| \mathbf{a}, \mathbf{b})$ in
version~1 (see Eq.~(\ref{eq:ProbBVer1}) above).

A model which violates parameter independence is one which would violate signal
locality in an imaginary world, where it was possible to obtain \emph{complete}
information regarding the state of the source, including the values of all 
the quantities which are in fact ``hidden''.  For version~2 of the Hess-Philipp
model  complete information regarding of the state of the source would have to
include  the value of the function $r(\lambda)$.

Suppose that we are in such an imaginary world.
As in the discussion following Eq.~(\ref{eq:ProbBVer1}) above, consider two
experimenters, Alice at $S_1$ and Bob at $S_2$.  Suppose that there is also a
third experimenter Xenophon located in the intersection of Alice and Bob's
backward light cones.  Xenophon prepares a succession of pairs in the singlet
state.  For each pair he determines the value of $r(\lambda)$ and then sends the
particles to Alice and Bob.  He also sends the value of $r(\lambda)$ to Bob (by
telephone, say).  Bob always  measures in the direction $(-1,0,0)$.  If Alice
measures in the direction $(1,0,0)$ then Bob obtains the value $r(\lambda)$ with
probability 1.  If, on the other hand, Alice measures in the direction $(0,1,0)$
there is only probability $1/2$ of Bob obtaining the value $r(\lambda)$.  The
value $r(\lambda)$ is known to Bob.  It is therefore possible for Alice to send
superluminal signals to Bob, just as in the case discussed in the passage
following Eq.~(\ref{eq:ProbBVer1}) above.

Finally, let us note that it does not help if one modifies the model  again, by
making
$r$ a function of the state of the detectors instead of $\lambda$. It is easily
verified that the model, thus modified, would  still violate
Shimony's~\cite{Shimony2,Shimony3} condition of outcome dependence
(``completeness'' in the terminology of Jarrett~\cite{Jarrett})---meaning that the
model would still be non-local.  Moreover, the model would still allow
Alice and Bob to communicate superluminally in an imaginary world were it was
possible to obtain complete information regarding the state of any system
(although the reading of Bob's detector is given by $r B$, in such a world Bob
could still find out the value of $B$ by inspecting the hidden state of his
detector).  This means that, on the level of the hidden variables, there must be
superluminal exchanges of information between the two detectors. 

\section{Conclusion}

We conclude that the expressions for the 
probabilities
$p_1$ and
$p_2$ calculated in this paper have the inescapable implication that the
Hess-Philipp model is non-local.  Hess and Philipp are only entitled to challenge
this conclusion if they can find an error in our calculations, and if they can
provide  alternative calculations leading to expressions for $p_1$ and $p_2$
which do not have the implication of non-locality.

\section*{Acknowledgment}
We are grateful to  M.~\.{Z}ukowski for some very useful
discussions, and for his  encouragement.

\appendix
\section{Remarks on the Combinatoric Part of the Construction}
In this appendix we first discuss a minor technical difficulty with the 
construction leading to Eq.~(35) of reference~\cite{Hess1}.  We then go on to
derive the expression for $E_{\lambda} (A_{\mathbf{a}})$
given in Eq.~(\ref{eq:ACondExp}) above.

Hess and Philipp choose in turn each of the $(n+1)^2$   squares $Q_{j
k}$.  For each choice of $Q_{jk}$ they define $L$ measures, where $L$ is the
binomial coefficient $\mathstrut^{9 n^2} \mspace{-6 mu} C_{3n}$. Each of these
$L$  measures assigns the weights $|a_1 b_1|$, $|a_2 b_2|$, $|a_3 b_3|$ to $3$
of the unit squares $\in Q_{j k}$, and the weights $(1/2) N_{1}
(|a_1|)
\psi_1 (|b_1|), \dots, (1/2) N_{n} (|a_3|) \psi_n (|b_3|)$ to $3 n$ of the $9 n^2$
unit squares not contained in  the vertical and horizontal strips defined by
$Q_{jk}$.  It assigns every other unit square  weight $0$.  The
assignment is such that each unit square
$\in Q_{jk}$ is assigned the weight $|a_t b_t|$ by $L/9$ measures, and each unit
square not in the vertical and horizontal strips defined by $Q_{j k}$ is assigned
the weight $(1/2) N_i(|a_t|) \psi_i(|b_t|)$ by $L/(9 n^2)$ measures ($t = 1,2,3$
and $i=1, \dots, n$).

The problem with this construction is that it tacitly assumes that the binomial
coefficient 
$\mathstrut^{9 n^2} \mspace{-6 mu} C_{3n}$ is divisible by $9
n^2$.  However, this is typically not the case (the only even integers $\le
100$ for which it is true  are $n= 10, 40, 44$ and  $84$).  Moreover, it is not
clear to us that the other requirements can be satisfied even when $n$ does
satisfy this condition.  The difficulty is, however, easily resolved if, instead
of taking $L=\mathstrut^{9 n^2} \mspace{-6 mu} C_{3n}$, we take it to be the
permutation $\mathstrut^{9 n^2} \mspace{-6 mu} P_{3n} = (9 n^2)!/ (9 n^2 - 3 n)!$.

Now let us turn to the derivation of Eq.~(\ref{eq:ACondExp}).  Let $I_{j k}$ be
the set of  indices $m$ such that $\mu_m$ is one of the $L$ measures
associated with $Q_{j k}$.  It is easily seen that
\begin{equation}
\sum_{m \in I_{j k}} \int_{Q_{jk}} A_m (u)\, d \mu_m = L \mathbf{a} \cdot
|\mathbf{b}|
\end{equation}
Let $S_{j k}$ be the subset of $[-3, 3n)^2$ which is obtained by deleting the
horizontal and vertical strips defined by $Q_{j k}$.  For each $m\in I_{j k}$,
$\int_{U} A_{m} (u) du dv =1$ for half the unit squares $U\subset S_{j k}$, and
it $=0$ for the other half.  Consequently
\begin{equation}
\sum_{m \in I_{j k}}\int_{S_{j k}} A_{m} (u) \, d\mu_m =  \frac{L}{4} \sum_{t =
3}^{3} \sum_{i=1}^{n} N(|a_{t}|) \psi(|b_{t}|)
= \frac{L}{2} \left( 1 - |\mathbf{a}| \cdot |\mathbf{b}| + \frac{\theta}{8 n^2}
\right) 
\end{equation}
with $0 \le \theta \le 1$ (where we have used Eq.~(21) of ref.~\cite{Hess1}).
Hence
\begin{equation}
\sum_{m \in I_{j k}} \int_{[-3,3n)^2} A_{m} (u) \, d\mu_m
= L \left(\mathbf{a} \cdot
|\mathbf{b}| +\frac{1}{2}( 1 - |\mathbf{a}| \cdot |\mathbf{b}|)
 + \frac{\theta}{16
n^2}
\right) 
\end{equation}
The right-hand side of this equation is independent of $j, k$.  Consequently, the
effect of summing over all $j$, $k$ is simply to multiply the expression by
$(n+1)^2$.  After dividing by  $N= (n+1)^2 L$ this gives
\begin{equation}
E_{\lambda} (A_{\mathbf{a}}) 
= \frac{1}{N} \sum_{m=1}^{N} \int_{[-3,3n)^2} A_{m} (u) \,
d\mu_m = \mathbf{a} \cdot
|\mathbf{b}| +\frac{1}{2}( 1 - |\mathbf{a}| \cdot |\mathbf{b}|)
 + \frac{\theta}{16
n^2}
\end{equation}

\end{document}